\documentclass[
  pre,
  aps,
  a4paper,
  english,
  showpacs,
  reprint,
  twocolumn,
  superscriptaddress,
]{revtex4-1}

\usepackage[T1]{fontenc}
\usepackage[utf8]{inputenc}
\usepackage{amsmath}
\usepackage{amssymb}
\usepackage{graphicx}
\usepackage{refstyle}
\usepackage[
  citecolor=blue,
  colorlinks,
  linkcolor=blue,
  urlcolor=blue,
]{hyperref}
\usepackage[dvipsnames]{xcolor}

\begin{document}

\title{Equivalent linearization finds nonzero frequency corrections beyond first order}
\author{Rohitashwa Chattopadhyay}
\email{crohit@iitk.ac.in}
\affiliation{
  Department of Physics,
  Indian Institute of Technology Kanpur,
  Uttar Pradesh 208016, India
}
%
\author{Sagar Chakraborty}
\email{sagarc@iitk.ac.in}
\affiliation{
  Department of Physics,
  Indian Institute of Technology Kanpur,
  Uttar Pradesh 208016, India
}
\affiliation{
  Mechanics and Applied Mathematics Group,
  Indian Institute of Technology Kanpur,
  Uttar Pradesh 208016, India
}
\begin{abstract}
We show that the equivalent linearization technique, when used properly, enables us to calculate frequency corrections of weakly nonlinear oscillators beyond the first order in nonlinearity. We illustrate the method by applying it to the conservative anharmonic oscillators and the nonconservative van der Pol oscillator that are respectively paradigmatic systems for modeling center-type oscillatory states and limit cycle type oscillatory states. The choice of these systems is also prompted by the fact that first order frequency corrections may vanish for both these types of oscillators, thereby rendering the calculation of the higher order corrections rather important. The method presented herein is very general in nature and, hence, in principle applicable to any arbitrary periodic oscillator.
\end{abstract}

%
%
\maketitle
\section{Introduction}
A linear autonomous (explicitly time independent) dynamical system is exactly solvable but even a weakly nonlinear term, when added to such system, makes the solutions of the system analytically elusive. In fact, a far-reaching discovery of the last century has been that, in three-or-more dimensional systems, nonlinearity introduces chaotic solutions which are not at all conceivable in linear systems. While the numerical studies of nonlinear systems are unavoidable and very fruitful, a plethora of useful analytical perturbative techniques~\cite{AHN2000} are available for such systems. Such techniques, apart from acting as benchmarks for numerical results, also provide deep insight into the nonlinear systems.  In this context, it is worthwhile to appreciate how the canonical perturbation theory, having very contemporary applications~\cite{JHJJJ2015PRL}, has remained a cornerstone in the centuries-old subject of classical mechanics. Actually, this perturbative technique remains an indispensable analytical tool because most realistic problems---classical three-body problem~\cite{CEN07} being probably the most celebrated one---are not exactly integrable. It should also be noted that the ideas behind the classical perturbation methods have been naturally extended to attack those problems in quantum mechanics that are not amenable to exact solutions in closed form. There is a very recent insightful exercise~\cite{BR2016AJP} on how perturbative renormalization group theory~\cite{CGO1996PRE} may be extended to tackle the quantum mechanical problems of the time dependent perturbation theory.

Another interesting example of the perturbative methods is the method of equivalent linearization of Kryloff and Bogoliuboff~\cite{MIN47,NWM50,C1963JASM}. Unlike canonical perturbation theory, it holds for dissipative systems as well. The essence of this technique is very simple and is used across disciplines, e.g., an analogue of this linearization idea, known as Hartee approximation~\cite{MAH20}, is utilised in approximating quantum many body systems to good effect. Also, the equivalent linearization coupled with the Ehrenfest theorem provides a way to analyze analytically intractable quantum dynamics of purely anharmonic oscillators~\cite{SB2015EJP}. The method of equivalent linearization is unique in the sense that in spite of being a perturbative technique when properly used it even yields results for the cases where the nonlinearity is very strong~\cite{BBM1984PRA}. Additionally, it vastly improves the standard Lindstedt--Poincar\'e technique by extending the range of nonlinear parameters over which the later may be applicable. However, any classical text~\cite{MIN47,JS07} would conclude that the equivalent linearization method finds only the nonlinearity induced first order correction to the frequency of the underlying linear oscillator. In what follows, we clarify what we mean by it.
\section{Equivalent linearization}
Consider the equation of an anharmonic oscillator given by:
\begin{equation}
  \ddot{x} + \omega^2 x+\varepsilon x^{2 \mu-1}=0;\quad\mu\in\{2,3,4,\cdots\}.\label{eq:qo}
\end{equation}
This is a Hamiltonian system wherein a particle of unit mass is moving under the influence of a potential $V(x)=\omega^2x^2/2+\varepsilon x^{2 \mu}/2 \mu$, $\varepsilon$ being a positive number. In order to perform the equivalent linearization, we write the above equation as,
\begin{equation}
\ddot{x} + (\omega^2 + \alpha\varepsilon \langle x^{2 \mu-2}\rangle)x = -\varepsilon(x^3 - \alpha\langle x^{2 \mu-2} \rangle x)\,,\label{eq:qoe}
\end{equation}
where $\alpha$ is an undetermined constant and the angular brackets denote the average over one cycle of the oscillations.
Naturally, the modified frequency $\Omega$ now is given by
\begin{equation}
  \label{eq:elomega}
 \Omega^2 = \omega^2 + \alpha\varepsilon\langle x^{2 \mu-2}\rangle\,,
 \end{equation}
with $\langle\cdots\rangle=:\int_0^{2\pi/\Omega}\cdots dt$, if the right hand side of Eq.~(\ref{eq:qoe}) happens to be negligible. The idea of the equivalent linearization is to make the dynamical equation as close possible to the equation of motion of simple harmonic oscillator. This means that the parameter $\alpha$ has to be chosen so that the right hand side of Eq.~(\ref{eq:qoe}) is zero.  While this is not possible entirely, we can definitely choose $\alpha$ so as to make the lowest (thus, most dominant) harmonic terms vanish. Incidentally, this also effectively gets rid of secular terms that cause unwanted resonance. Under the condition of weak nonlinearity, mathematically manifested by small enough $\varepsilon$, it is reasonable to assume the solution of the form
\begin{equation}x = a\cos{\Omega t}\label{eq:ansatz}\end{equation}
corresponding to the initial conditions: $x(0) = a, \dot{x}(0) = 0$; which are chosen for the sake of simplicity without any loss of generality. Now, the choice $\alpha \langle x^{2 \mu-2}\rangle= \frac{a^{2\mu-2}}{\mu 2^{2\mu-1}}\frac{\Gamma\left(2\mu+1\right)}{\Gamma\left(\mu+1\right)\Gamma\left(\mu\right)}$ gets rid $\cos\Omega t$ term in the right hand side of Eq.~(\ref{eq:qoe}). Here $\Gamma$ denotes the gamma function. We can now write Eq.~(\ref{eq:elomega}) as,
\begin{equation}
  \label{eq:elnewomega}
  \Omega^2 = \omega^2 +\varepsilon\frac{a^{2\mu-2}}{\mu 2^{2\mu-1}}\frac{\Gamma\left(2\mu+1\right)}{\Gamma\left(\mu+1\right)\Gamma\left(\mu\right)}.
\end{equation}
Therefore, we note that \emph{the equivalent linearization finds the first order frequency correction}. It may be noted that although, in view of Eqs.~(\ref{eq:qoe}) and (\ref{eq:ansatz}), this method is similar in spirit to the `iteration method' (an improved version of Duffing's idea~\cite{JS1950}), but the method of equivalent linearization uses an inherently different idea---the idea of averaging---which is a source of problems as discussed below.
\section{The problems and their resolutions}
The preceding sequence of mathematical steps (Eqs.~(\ref{eq:qo})-(\ref{eq:elnewomega})) seems very general. However, it hits a stumbling block as soon as we merely replace $x^{2\mu-1}$ in Eq.~(\ref{eq:qo}) by $x^{2\mu}$, with $\mu\in\{1,2,3,\cdots\}$. Lets us elaborate on it in what follows. We have the equation of motion of an anharmonic oscillator: 
\begin{equation}
\label{cubic}
\ddot{x}+x+\varepsilon x^{2\mu} = 0;\quad \mu\in\{1,2,3,\cdots\}.
\end{equation}
Here without any loss of generality, we have put $\omega=1$. This again is a Hamiltonian system in which a particle of unit mass moves under the effect of a potential $V(x)=x^2/2+\varepsilon x^{2\mu+1}/(2\mu+1)$. For small enough initial amplitude, the particle can oscillate.
We rewrite Eq.~(\ref{cubic}) as,
\begin{equation}
\label{cubicel}
\ddot{x}+\Omega^2x = -\varepsilon x^{2\mu}+\alpha \varepsilon \langle x^{2\mu-1}\rangle x\,,
\end{equation}
where,
\begin{equation}
  \label{eq:elomegac}
 \Omega^2 = 1 + \alpha\varepsilon\langle x^{2\mu-1}\rangle\,.
\end{equation}
Lets choose the trial oscillatory solution as,
\begin{equation}
x=a \cos \Omega t\,.\label{eq:xacc}
\end{equation}
We immediately note the \emph{first problem} that $\langle x^{2\mu-1} \rangle$, i.e., $x$ averaged over the duration between $0$ to $2 \pi/\Omega$ yields zero rendering $\alpha$ indeterminate. \emph{We propose to overcome this problem by redefining the average as}: $\langle\cdots\rangle=:\int_0^{2\pi}\cdots dt$, i.e., averaging is being performed over the unperturbed time period (time period of the linear system $\ddot{x}+x=0$). Now the dominant harmonic term in the right hand side of Eq.~(\ref{cubicel}) vanishes only if we set $\alpha =0$. Hence, the first order correction to frequency is zero [please see Eq.~(\ref{eq:elomegac})], or mathematically,
\begin{equation}
\Omega^2=1+0\varepsilon+\mathcal{O}(\varepsilon^2)\,.\label{eq:newomegac}
\end{equation}
One could argue that under the assumption of finite $\alpha$, the older definition of $\langle\cdots\rangle$ is sufficient in getting us the first order vanishing frequency correction. Nevertheless, apart from the fact that having undetermined $\alpha$ is not so satisfactory, we shall discover shortly that the older definition is not effective in evaluating the nonzero frequency corrections in other types of oscillators, e.g., van der Pol oscillator. 

While Eq.~(\ref{eq:newomegac}) is the correct result, the \emph{second problem} one could subsequently pose is: what is the first nonzero frequency correction? Is its calculation in the ambit of the equivalent linearization method? To the best of our knowledge this important question has not been explored in the rich literature of perturbation theories. In fact, in a bit stronger words, one can say that it is a misconception that the equivalent linearization is valuable only in detecting the first order frequency corrections. We now show \emph{the main result of this paper that the equivalent linearization technique helps to find the nonzero frequency corrections even if they are of second or higher order}.

To this end, we make use of Eq.~(\ref{eq:xacc}) in the right hand side of Eq.~(\ref{cubicel}) and set $\alpha=0$. Thus, we arrive at the following differential equation:
\begin{equation}
\label{redcu}
\ddot x+\Omega^2 x=-\varepsilon a^{2\mu} \cos^{2\mu} \Omega t\,.
\end{equation}
This is easily integrated to find
\begin{equation}
\label{amplitudecorr}
x=a\cos\Omega t-\frac{\varepsilon a^{2\mu}}{\Omega^2}\left[\sum^{\mu}_{m=1}\frac{c_m \cos\left(2m\Omega t\right)}{\left(2m\right)^2-1}+\frac{c_0}{2}\right],
\end{equation}
where, $c_n=\frac{2}{2^{2\mu}\left(2\mu+1\right) \beta\left(\mu+n+1,\mu-n+1\right)}$
for $n\in\{0,1,2,\cdots,\mu\}$.
$\beta$ denotes the beta function, also known as the Euler integral of the first kind.
We remark that in order to obtain $c_n$, one needs to evaluate integrals~\cite{GR1980} of the form $\int_0^{2\pi} \cos^n x \cos mx$ ($n, m \in \{1,2,\cdots\}$).
Now instead of using $x$ from Eq.~(\ref{eq:xacc}), we use its expression as given by Eq.~(\ref{amplitudecorr}) in Eq.~(\ref{cubicel}). With a view to suppressing the lowest harmonic terms, we invariably need, 
\begin{equation}
\alpha \langle x^{2 \mu-1}\rangle= \varepsilon\left(\frac{2\mu a^{4\mu-2}}{\Omega^2}\right) \left(\sum^{\mu}_{m=1}\frac{c^2_m}{\left(2m\right)^2-1}-\frac{c_0 d_0}{2}\right),
\end{equation}
where $d_0$ has been defined as $\frac{2}{2^{2\mu-1} \left(2\mu\right) \beta\left(\mu+1,\mu\right)}$ and it should be kept in mind that angular brackets have the aforementioned redefined meaning.  Therefore, from Eq.~(\ref{eq:elomegac}), we get
\begin{equation}
\label{eq:expeq2}
\Omega^2=1+ \varepsilon^2 \left(\frac{2\mu a^{4\mu-2}}{\Omega^2}\right)\left(\sum^{\mu}_{m=1}\frac{c^2_m}{\left(2m\right)^2-1}-\frac{c_0 d_0}{2}\right)+\mathcal{O}\left(\varepsilon^3\right)
\end{equation}
which is the correct answer that matches with the ones obtained using alternate perturbative methods like Lindtsted--Poincar\'e technique etc. or exact method of evaluating the time period. For convenience, in Table~\ref{tab:1} we have illustrated the frequency corrections for some specific values of $\mu$.
%
\begin{table}
\caption{Explicit values of the frequency corrections, as given by Eq.~(\ref{eq:expeq2}) and Eq.~(\ref{eq:expeq1}), for anharmonic oscillators. For oscillators following equation of motion: $\ddot{x}+x+\varepsilon x^n=0$ ($n=2,3,4,\cdots,11$), we represent the frequency $\Omega$ as $1+\varepsilon \omega_1 + \varepsilon^2 \omega_2+\mathcal{O}\left(\varepsilon^3\right)$ whenever oscillatory solutions are possible for a given initial amplitude $a$ at which the oscillator is at rest.}
\label{tab:1} 
\begin{tabular}{ccc}
\hline\noalign{\smallskip}
n & 1$^{\rm st}$ order correction, $\omega_1$ & 2$^{\rm nd}$ order correction, $\omega_2$  \\
\noalign{\smallskip}\hline\noalign{\smallskip}
2 & 0 & $-\frac{5a^2}{12}$ \\
3 & $+\frac{3a^2}{8}$ & $-\frac{15a^4}{256}$\\
4 & 0 & $-\frac{63a^6}{160}$  \\
5 & $+\frac{5a^4}{16}$ & $-\frac{55a^8}{3072}$ \\
6 & 0 & $-\frac{302a^{10}}{841}$  \\
7 & $+\frac{35a^6}{128}$ &  $+\frac{758a^{12}}{66015}$ \\ 
8 & 0 &   $-\frac{396a^{14}}{1201}$  \\
9 & $+\frac{63a^8}{256}$ &  $+\frac{131a^{16}}{3841}$\\ 
10 & 0 & $-\frac{881a^{18}}{2881}$   \\
11 & $+\frac{924a^{10}}{4096}$ &  $+\frac{296a^{20}}{7601}$   
 \\
\noalign{\smallskip}\hline
\end{tabular}
\end{table}

In fact, one can even apply this procedure to Eq.~(\ref{eq:qo}) and find the second order correction to frequency as follows. Using Eq.~(\ref{eq:elomega}) in  Eq.~(\ref{eq:qo}) with the value of $\alpha\langle x^{2 \mu-2}\rangle= \frac{a^{2\mu-2}}{\mu 2^{2\mu-1}}\frac{\Gamma\left(2\mu+1\right)}{\Gamma\left(\mu+1\right)\Gamma\left(\mu\right)}$, we obtain the following differential equation:
\begin{equation}
\label{genquartic}
\ddot x+\Omega^2 x=-\varepsilon a^{2\mu-1} \cos^{2\mu-1} \Omega t\,,
\end{equation} 
which can also be integrated to find
\begin{equation}
\label{amplitudecorr1}
x=a\cos\Omega t-\frac{\varepsilon a^{2\mu-1}}{\Omega^2}\left[\sum^{\mu}_{m=2}\frac{c_m \cos\left\{\left(2m-1\right)\Omega t\right\}}{\left(2m-1\right)^2-1}\right]\,,
\end{equation}
where, $c_m=\frac{2}{2^{2\mu-1}2 \mu \beta\left(\mu+m,\mu-m+1\right)}.$ Again, using $x$ from Eq.~(\ref{amplitudecorr1}) in Eq.~(\ref{eq:qo}), we get
\begin{eqnarray}
\label{eq:expeq1}
&&\Omega^2=1+\varepsilon\left(\frac{a^{2\mu-2}}{\mu 2^{2\mu-1}}\right)\frac{\Gamma\left(2\mu+1\right)}{\Gamma\left(\mu+1\right)\Gamma\left(\mu\right)}\nonumber\\&&\qquad-\varepsilon^2\left[\frac{\left(2\mu-1\right) a^{4\mu-4}}{\Omega^2}\right]\sum_{m=2}^{\mu}\frac{c^2_m}{\left(2m-1\right)^2-1}+\mathcal{O}\left(\varepsilon^3\right).\nonumber\\
\end{eqnarray}
Table~\ref{tab:1} tabulates the corrections explicitly for some values of $\mu$.

As an application of the above results, it is easy to see that the above discussion can be, in principle, extended to analyse motion of a harmonic oscillator under the influence of an additional exponential potential of the form $V\left(x\right)=-\varepsilon\exp\left(-{x^2}/{b^2}\right)$---the constant $b^2$ being a positive number---that yields the following equation of motion,
\begin{equation}
\label{expeq}
\ddot x +x=-\varepsilon x\exp\left(-\frac{x^2}{b^2}\right).
\end{equation}
We rewrite Eq~(\ref{expeq}) as,
\begin{equation}
\label{expeq2}
\ddot x +\Omega^2 x=-\varepsilon x\exp\left(-{x^2}/{b^2}\right)+\varepsilon \alpha\left \langle\exp\left(-\frac{x^2}{b^2}\right)\right\rangle x,
\end{equation}
where, $\Omega^2=1+\varepsilon \alpha \langle\exp\left(-\frac{x^2}{b^2}\right)\rangle.$
Now, on expanding the exponential term in the right hand side of Eq.~(\ref{expeq2}), we get,
\begin{eqnarray}
\label{expeq2}
\ddot x +\Omega^2 x=\sum^{\infty}_{n=0}\left(
-\varepsilon_n{x^{2n+1}}+\varepsilon_n \alpha \left\langle{x^{2n}}\right\rangle x\right),
\end{eqnarray}
Here, $\varepsilon_n=\varepsilon(-1)^n/(b^{2n}n!)$. 
Note that Eq.~(\ref{expeq2}) looks like Eq.~(\ref{eq:qoe}) with its right hand side replaced by an infinite series. Hence, the frequency correction of Eq.~(\ref{expeq}) may be obtained as an infinite sum of terms each found following the same process outlined while finding the frequency correction for Eq.~(\ref{eq:qo}). However, a closed form of the correction at every order may not be possible.
\section{Frequency of van der Pol oscillator}
Now we present the interesting example of the van der Pol oscillator~\cite{K2007} to illustrate how our scheme of the equivalent linearization works even for limit cycle oscillations that are possible only in a nonconservative system for which usually Hamiltonian formalism is elusive. Although the method of equivalent linearization is frequently used to find approximate linearized equation of the trajectories near limit cycles~\cite{JS07}, it is seldom used to find the nontrivial frequency corrections for the unique isolated oscillatory state.  Let us witness how easily the equivalent linearization can accomplish the job for us. For this purpose, we work with the van der Pol oscillator, a well known dissipative system that possess a stable limit cycle and is widely used pedagogical model of limit cycle oscillations. The equation of motion for the van der Pol oscillator is 
\begin{equation}
\ddot x+\varepsilon \left(x^2-1\right)\dot x+x=0\,,
\end{equation}
which can be recast as
\begin{equation}
\label{vdpoel}
\ddot x+\Omega^2 x=-\varepsilon \left(x^2-1\right)\dot x+\varepsilon\alpha\langle x \dot x\rangle x\,.
\end{equation}
As before we define $\Omega^2=1+\varepsilon\alpha\langle x \dot x \rangle$.
We again note that in order to have a non-zero value for $\langle x \dot x\rangle$, $x\dot{x}$ must averaged using the limits $0$ and $2\pi$ (and not $2\pi/\Omega$). Substituting $x=a \cos \Omega t$ in the right hand side of Eq.~(\ref{vdpoel}), we find that in order to set lowest harmonic term involving $\cos\Omega t$ to zero, we require $\alpha =0$. Consequently we note that the first order correction to frequency is zero in accordance with the definition of $\Omega$. Additionally, setting the coefficient of $\sin \Omega t$ to zero, we get the amplitude of the limit cycle as $a=2$. Thus, Eq.~(\ref{vdpoel}) simplifies to
\begin{equation}
\ddot x+\Omega^2 x=\varepsilon \frac{2^3}{4}\Omega \sin 3\Omega t\,,
\end{equation}
that describes the evolution of a phase point on the limit cycle of radius $2$.
Solving the above equation gives the first order correction for the solution for the limit cycle:
\begin{equation}
\label{vpdoamplitudecorr}
x=2\cos \Omega t-\varepsilon \frac{2^3}{32 \Omega}\sin 3\Omega t\,.
\end{equation}
As outlined for the case of anharmonic oscillators, we substitute Eq.~(\ref{vpdoamplitudecorr}) in the right hand side of Eq.~(\ref{vdpoel}) and set the coefficient of the most dominant harmonic to zero. Consequently, $\alpha=-\varepsilon {2^4}/{128 \langle x\dot x\rangle}$ which in turn implies that
\begin{equation}
\Omega=1+0\varepsilon-\frac{1}{16}\varepsilon^2 +\mathcal{O}\left(\varepsilon^3\right)\,.
\end{equation}
This again is a well-known correct expression~\cite{SCVC2015PRE,SCC2016ARXIV}.
\section{Conclusion}
Summarizing, we have shown that the method of the equivalent linearization is much more powerful a method than traditionally realized. It can be used to calculate frequency corrections of weakly nonlinear oscillators to arbitrary orders. However, one must note that for certain cases where the average $\langle\cdots\rangle$ vanishes, one must redefine it so that it stands for average over the unperturbed time period. This improvisation extends the scope of the equivalent linearization immensely to include systems like the cubic anharmonic oscillator and the van der Pol oscillator. Thus, equivalent linearization can be employed beyond first order in both the conservative and the nonconservative systems. One may also note that the equivalent linearization doesn't require any prior ansatz for the perturbative series of the dynamical variable unlike several other methods, e.g.,  canonical perturbation theory, coordinate perturbation method, Lie transform perturbation theory, Lindtsted--Poincar\'e method, multiple time scale method, renormalization group method, etc.~\cite{AHN2000,MIN47,JS07,SCVC2015PRE,SCC2016ARXIV,BMC2007}. In fact, the perturbative series for the solution is automatically generated order by order as one keeps on extracting the frequency corrections order by order.
The authors are grateful to Jayanta Kumar Bhattacharjee, Anindya Chatterjee, and Tirth Shah for fruitful discussions.


\begin{thebibliography}{13}
\bibitem{AHN2000} A. H. Nayfeh, \textit{Perturbation Methods} (Wiley-{VCH} Verlag {GmbH}, Berlin, 2000).
%
\bibitem{JHJJJ2015PRL} D-O Jeon, K. R. Huang, J-H Jang, H. Jin, and H. Jang, Phys. Rev. Lett. \textbf{114},  184802 (2015).  
%
\bibitem{CEN07} A. Chenciner, Scholarpedia \textbf{2},  2111 (2007). 
%
\bibitem{KZ2004AJP} P. B. Kahn and Y. Zarmi, Am. J. Phys. \textbf{72}, 538 (2004).
%
\bibitem{BR2016AJP} J. K. Bhattacharjee and D. S. Ray, Am. J. Phys. \textbf{84}, 434 (2016).
%
\bibitem{CGO1996PRE} L-Y Chen,  N. Goldenfeld,  and Y. Oono, Phys. Rev. E \textbf{54}, 376 (1996).
%
\bibitem{TDF1992} C. Cohen-Tannoudji, B. Diu, F. Laloe, \textit{Quantum Mechanics (2 vol. set)} (Wiley-{VCH} Verlag {GmbH}, Berlin, 1992). 
%
\bibitem{G2005} D. J. Griffiths, \textit{Introduction to Quantum Mechanics} (Pearson Prentice Hall, Upper Saddle River, NJ, 2005).
%
\bibitem{S1994} R. Shankar, \textit{Principles of Quantum Mechanics} (Springer, New York, 1994).
%
\bibitem{MIN47} N. Minorsky, \textit{Introduction to Nonlinear Mechanics: Topological Methods, Analytical Methods, Nonlinear Resonance, Relaxation Oscillations} (J.W. Edwards, Ann Arbor, Michigan, 1947). 
%
\bibitem{NWM50} N. W. McLachlan, \textit{Ordinary Nonlinear Differential Equations in Engineering and Physical Sciences} (Clarendon Press, Oxford, 1950). 
%
\bibitem{GR1980} I. S. Gradshteyn and I. M. Ryzhik,  \textit{Table of Integrals, Series, and Products (Corrected and Enlarged Edition)}  (Academic Press, 1980). 


%
\bibitem{C1963JASM} T. K. Caughey, J. Acoust. Soc. Am. \textbf{35},  1706 (1963).
%
\bibitem{MAH20} G. D. Mahan, \textit{Many-Particle Physics (Physics of Solids and Liquids)} (Springer, New York, 2000).  
%
\bibitem{SB2015EJP} R. Sain and J. K. Bhattacharjee, Eur. J. Phys. \textbf{36}, 055025 (2015).
%
\bibitem{BBM1984PRA} K. Banerjee, J. K. Bhattacharjee, and H. S. Mani, Phys. Rev. A \textbf{30}, 1118 (1984).
%

\bibitem{JS07} D. W. Jordan and P. Smith, \textit{Nonlinear Ordinary Differential Equations: Problems and Solutions: A Sourcebook for Scientists and Engineers (Oxford Texts in Applied and Engineering Mathematics)} (Oxford University Press, Oxford, 2007).
%
\bibitem{JS1950} J. J. Stoker, \textit{Nonlinear Vibrations in Mechanical and Electrical Systems} (Wiley-Interscience, New York, 1950).
%
\bibitem{K2007} T. Kanamaru, Scholarpedia \textbf{2},  2202 (2007). 
%
\bibitem{SCVC2015PRE} T. Shah, R. Chattopadhyay, K. Vaidya, and S. Chakraborty, Phys. Rev. E \textbf{92}, 062927 (2015).
%
\bibitem{SCC2016ARXIV} T. Shah, R. Chattopadhyay, and S. Chakraborty, 	arXiv:1610.05218 (2016).
%
\bibitem{BMC2007} J. K. Bhattacharjee, A. K. Mallik, and S. Chakraborty, Indian J. Phys. \textbf{81}, 1115 (2007).



\end{thebibliography}
\end{document}